\documentclass[%
twocolumn,
superscriptaddress,
preprintnumbers,
amsmath,amssymb,
pra,
]{revtex4-2}

\usepackage{amsmath}
\usepackage{amssymb}
\usepackage{graphicx}
\usepackage{color}
\usepackage[colorlinks=true, linkcolor=blue, citecolor=blue, urlcolor=blue]{hyperref}
\usepackage{epsfig}
\usepackage{amsmath}
\usepackage{amssymb}
\usepackage{epstopdf}
\usepackage{sidecap}
\usepackage{booktabs}
\usepackage{rotating}
\usepackage[bottom]{footmisc}
\usepackage{quantikz} 

\sidecaptionvpos{figure}{c}

\DeclareGraphicsExtensions{.png .jpg .pdf}

\renewcommand{\j}{\mathrm{i}}
\newcommand{\e}{\mathrm{e}}

\newcommand{\mc}{\mathcal}

\newcommand{\mr}{\mathrm}

\begin{document}

\title{Performance and Stability of Quantum Krylov Diagonalization for the Hubbard Model}

\author{Mohammad Mirzakhani}\email{mirzakhani@sejong.ac.kr}
\affiliation{Department of Physics and Astronomy, Sejong University, Seoul 05006, South Korea}
\affiliation{Department of Physics, Yonsei University, Seoul 03722, South Korea}

\author{Hyukgun Kwon}\email{kwon37hg@sejong.ac.kr}
\affiliation{Department of Physics and Astronomy, Sejong University, Seoul 05006, South Korea}

\author{Kyungsun Moon}\email{kmoon@yonsei.ac.kr}
\affiliation{Department of Physics, Yonsei University, Seoul 03722, South Korea}

\date{March 09, 2026}
\begin{abstract}
    Quantum Krylov diagonalization (QKD) has emerged as a promising hybrid quantum-classical approach 
    for estimating ground-state properties of many-body systems on near-term quantum devices. 
    In this work, we investigate the convergence, stability, and hardware performance of QKD for 
    the one-dimensional Hubbard model with periodic boundary conditions. 
    Building upon our previously developed low-depth Jordan--Wigner implementation, which 
    reduces the number of two-qubit (CNOT) gates required for quantum time evolution, we perform a 
    systematic study of the influence of the Krylov dimension, Hamiltonian evolution parameters, 
    system size, interaction strength, and singular-value truncation (SVT) on the convergence 
    of the method. 
    Our results show that the performance of QKD is governed by a delicate interplay between 
    the low-energy spectral structure of the Hamiltonian and numerical stability. 
    In particular, systems with near-closing energy gaps require longer evolution times to 
    efficiently resolve nearby eigenstates, while the evolution time, Krylov dimension, 
    Trotter number, and SVT threshold must be carefully balanced to avoid numerical 
    instabilities and accumulated time-discretization errors. 
    This analysis provides practical guidelines for selecting algorithmic parameters in QKD. 
    Finally, we demonstrate the algorithm on IBM quantum hardware, where the experimental results 
    reproduce the convergence trends predicted by ideal simulations using only lightweight 
    readout-error mitigation and a modest measurement budget. 
    Together, these results demonstrate that QKD is a practical and hardware-efficient approach 
    for studying strongly correlated fermionic systems on current NISQ quantum processors.
\end{abstract}

\maketitle

\section{Introduction}  \label{sec:intro}
Understanding the ground-state properties of strongly correlated quantum systems remains one 
of the central challenges in condensed matter physics owing to the exponential growth of 
the many-body Hilbert space with system size~\cite{Feynman1982,Lloyd1996,Bloch2008}.
Among the various models describing correlated electrons, the Hubbard model \cite{Hubbard1963} 
occupies a central role as it captures the essential competition between electron hopping 
and on-site Coulomb 
interactions, giving rise to a wide range of quantum phenomena, including Mott 
insulating behavior, quantum magnetism, and unconventional
superconductivity~\cite{Hubbard1963,LiebWu1968,Essler2005,Qin2022}. 
Although classical numerical methods, such as exact diagonalization, provide 
numerically exact solutions for finite systems, their computational cost grows 
exponentially with system size, severely restricting their applicability to relatively 
small lattices \cite{Feynman1982,Orus2014}. 
Quantum computing offers a promising alternative by exploiting the principles 
of quantum mechanics to efficiently represent and manipulate many-body quantum 
states \cite{Feynman1982,Lloyd1996, Georgescu2014,Bauer2016,Aspuru2005}. 
However, realizing this potential on current noisy intermediate-scale quantum (NISQ) 
devices remains a significant challenge, as many quantum algorithms require deep 
circuits that exceed the coherence times and gate fidelities of existing hardware \cite{Kitaev1995,Abrams1997,Nielsen2000,Ortiz2001,Whitfield2011,Wecker2015solving,
    Babbush2018,Preskill2018,McArdle2020}.

Variational quantum algorithms have emerged as one of the leading approaches for estimating 
ground-state properties on NISQ devices~\cite{Peruzzo2014,McClean2016,Kandala2017,
    Cerezo2021,Stanisic2022}. 
Among them, the variational quantum eigensolver (VQE) employs a hybrid quantum-classical framework 
in which a parameterized trial state is optimized using a classical optimizer while expectation 
values of the Hamiltonian are evaluated on a quantum processor~\cite{Peruzzo2014,McClean2016,Cerezo2021}. 
Owing to its flexibility, VQE has been successfully applied to a wide range of problems 
in quantum chemistry and condensed matter physics, and numerous ans\"{a}tze and extensions have 
been proposed to improve its accuracy and 
applicability~\cite{Szalay1995,Harsha2018,Peruzzo2014,Dallaire2019,McClean2016,Kandala2017,Grimsley2019,
    McClean2017,Higgott2019,Santagati2018,Wang2019}.

Despite these advances, VQE faces several practical challenges. 
The repeated quantum-classical optimization cycle requires numerous circuit 
evaluations, resulting in significant measurement overhead. 
Moreover, the optimization landscape may become increasingly difficult to 
navigate because of stochastic measurement noise, ansatz limitations, barren 
plateaus, and local minima, all of which can hinder convergence and limit
scalability~\cite{Wecker2015,Gonthier2022}. 
These challenges have motivated the development of alternative hybrid 
quantum-classical approaches that avoid iterative optimization while 
remaining compatible with near-term quantum hardware.

Quantum subspace methods have recently emerged as a promising alternative to variational 
approaches for estimating eigenvalues of many-body Hamiltonians on near-term quantum
devices~\cite{Parrish2019,Motta2020}. 
Among these methods, quantum Krylov diagonalization (QKD) has attracted considerable 
attention owing to its ability to combine quantum time evolution with classical subspace
diagonalization~\cite{Saad2011,Parrish2019,Motta2020,Stair2020,Cohn2021,Seki2021,
    Epperly2022,Klymko2022,Cortes2022,Kirby2023Exact,Lee2024,Baker2024,Kirby2024Ana,
    Yoshioka2025,Oliveira2025}. 
Starting from a suitable reference state, QKD constructs a Krylov subspace by 
repeatedly applying the time-evolution operator, while the quantum processor 
prepares the Krylov basis states and evaluates the Hamiltonian and overlap matrix 
elements. 
The resulting generalized eigenvalue problem is then solved efficiently on a 
classical computer, combining the strengths of quantum state preparation with 
well-established classical linear algebra techniques. Owing to the absence of an 
iterative quantum-classical optimization loop, QKD offers an attractive framework 
for NISQ devices. Its practical feasibility has recently been demonstrated experimentally 
on superconducting quantum processors~\cite{Yoshioka2025}.

Despite these encouraging developments, a comprehensive understanding of the convergence and 
stability of QKD for strongly correlated fermionic systems remains lacking. 
In particular, the Hubbard model~\cite{Hubbard1963} presents several challenges, 
including nonlocal fermion-to-qubit mappings such as the Jordan--Wigner (JW) and Bravyi--Kitaev
transformations~\cite{Jordan1928,Bravyi2002,Seeley2012}, the large number of Pauli 
operators appearing in the qubit Hamiltonian~\cite{Whitfield2011,Babbush2018}, 
and numerical instabilities associated with the generalized eigenvalue problem arising 
from noise and the near-linear dependence of Krylov basis states~\cite{Stair2020,Klymko2022,
    Cortes2022,Baker2024}. 
Furthermore, the influence of key algorithmic parameters, including the Krylov dimension,
time-evolution parameter, Trotterization, system size, interaction strength, and 
singular-value truncation (SVT), has not yet been systematically investigated. 
Understanding how these factors govern the accuracy, stability, and experimental cost 
of QKD is essential for its reliable implementation on current quantum hardware.

In this work, we present a systematic investigation of the QKD algorithm for
computing the ground-state energy (GSE) of the half-filled one-dimensional (1D) 
Hubbard model with periodic boundary conditions (PBC). 
Building upon our previously developed low-depth circuit implementation 
\cite{Mirzakhani2025}, we analyze the convergence and stability of QKD over a broad 
range of algorithmic and physical parameters, including the Krylov dimension, 
time-evolution parameter, Trotter number, system size, interaction strength, and 
SVT threshold. 
Our study shows that the performance of QKD is governed by a
delicate interplay between the low-energy spectral structure of the Hamiltonian
and numerical stability. In particular, we demonstrate that systems with
near-closing energy gaps require longer evolution times to efficiently resolve
nearby eigenstates, while the evolution time, Krylov dimension, Trotter number,
and SVT threshold must be carefully balanced to avoid
numerical instabilities and accumulated time-discretization errors. 
Finally, we validate the proposed approach on IBM quantum hardware, demonstrating 
that the experimentally measured ground-state energies reproduce the 
convergence trends predicted by ideal simulations using only lightweight 
readout-error mitigation. 
These results provide practical guidelines for selecting the key
algorithmic parameters in QKD and further establish the method as a practical
and hardware-efficient approach for studying strongly correlated fermionic
systems on current NISQ quantum processors.

The remainder of this paper is organized as follows. 
Section~\ref{sec:Alg_Meth} introduces the Hubbard model and the QKD methodology, together with the quantum-circuit implementation.
Section~\ref{sec:results} presents a systematic analysis of the convergence and stability of the QKD algorithm under various physical and algorithmic parameters, followed by its experimental demonstration on IBM quantum hardware.
Finally, Sec.~\ref{sec:conc} summarizes the main findings and outlines possible directions for future work.

\section{Algorithm and methodology}  \label{sec:Alg_Meth}

\subsection{Hubbard Model and Quantum Circuit Implementation} \label{ss:QC_Hubbard}

The Hubbard model provides a minimal yet nontrivial description of
interacting electrons on a lattice, capturing the competition between
kinetic energy and on-site Coulomb repulsion \cite{Hubbard1963, Essler2005,Qin2022}. 
For a 1D chain with PBC, the Hubbard Hamiltonian is given by
\begin{align} \label{eq:Hubb}
    \mc {H} &= \mc{H}_0 + \mc{H_U} \notag \\
    &= -\gamma_0 \sum_{\langle i,j\rangle,\sigma}
    \left(c^\dagger_{i\sigma} c_{j\sigma} + c^\dagger_{j\sigma} c_{i\sigma}\right)
    + \mc U_0 \sum_i n_{i\uparrow} n_{i\downarrow},
\end{align}
where $c^\dagger_{i\sigma}$ ($c_{i\sigma}$) is the fermionic creation
(annihilation) operator for an electron with spin
$\sigma \in \{\uparrow,\downarrow\}$ at lattice site $i$, and
$n_{i\sigma} = c^\dagger_{i\sigma} c_{i\sigma}$ is the corresponding
number operator. 
The first term $\mc{H}_0$ describes the kinetic energy associated
with electron hopping between nearest-neighbor sites $\langle i,j\rangle$
with amplitude $\gamma_0$, while the second term $\mc{H_U}$ accounts for the on-site
Coulomb interaction of strength $\mc U_0 > 0$ between electrons of opposite spin
occupying the same lattice site.
Throughout this work we consider the half-filled case, $N_e=L$,
which corresponds to one electron per lattice site on average.

\begin{figure}
    \centering
    \includegraphics[width=0.9\linewidth]{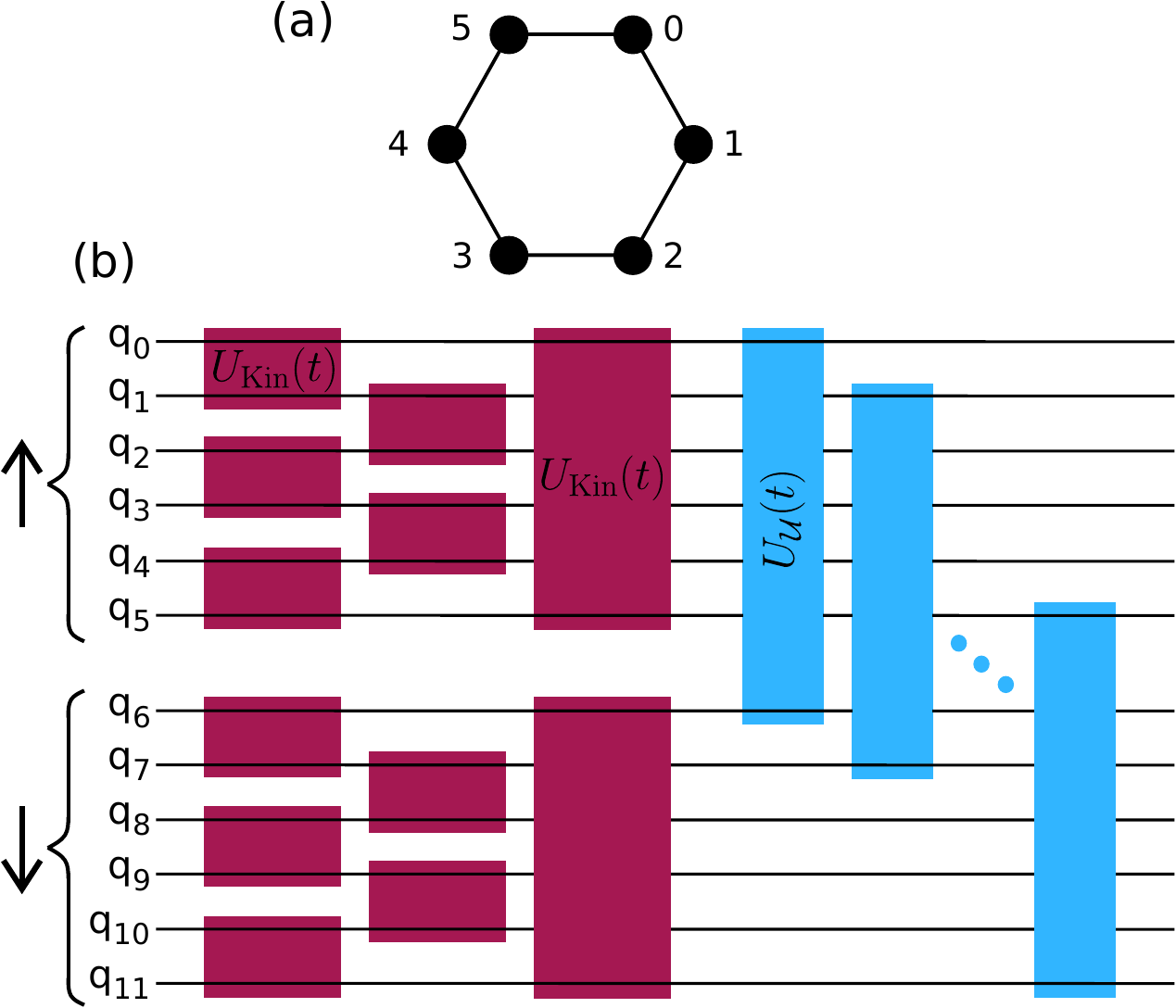}
    \caption{(a) 1D Hubbard chain with PBC,
        illustrated for the representative case of $L=6$ lattice sites labeled
        from $0$ to $5$.
        (b) Quantum circuit implementing the time-evolution operator of the
        Hubbard Hamiltonian [Eq.~\eqref{eq:Hubb}].
        The first $L$ qubits encode the spin-up orbitals, while the remaining
        $L$ qubits encode the spin-down orbitals.
        The red blocks, $U_{\mr{Kin}}(t)$, implement the hopping evolution
        between neighboring lattice sites, whereas the blue blocks,
        $U_{\mc U}(t)$, implement the on-site Coulomb interaction between the
        spin-up and spin-down qubits associated with the same lattice site
        ($q_i$ and $q_{L+i}$).
        Details of the circuit decomposition for
        $U_{\mr{Kin}}(t)$ and $U_{\mc U}(t)$ are given in
        Ref.~\cite{Mirzakhani2025}.
    } \label{fig1}
\end{figure}

The hopping term delocalizes electrons across the lattice and favors
metallic behavior, whereas the interaction term penalizes double
occupancy and promotes localization, leading to strongly correlated
phases. 
The interplay between these two competing energy scales gives
rise to a rich variety of physical phenomena, including Mott insulating
behavior \cite{Mott1949}, antiferromagnetic ordering, and correlation-driven phase
transitions. 
In the present work, we consider a chain of $L$ lattice
sites with PBC, such that a site $L$ is identified with site $0$, 
see Fig.~\ref{fig1}(a).
The total number of qubits required to represent the system is $2L$,
accounting for the two spin degrees of freedom per site as shown in Fig.~\ref{fig1}(b). 

To simulate the Hubbard model on a quantum computer, the fermionic operators are mapped to qubit 
operators using the JW transformation \cite{Ortiz2001,Whitfield2011,Bravyi2002,Jordan1928},
\begin{equation} \label{eq:JWformula}
    c_j = \left(\otimes_{k=1}^{j-1} Z_k \right) \otimes \sigma_j, \quad 
    c_j^\dagger = \left(\otimes_{k=1}^{j-1} Z_k \right) \otimes \sigma_j^\dagger,
\end{equation}
where $\sigma_j = \frac{1}{2} (X_j + \j Y_j)$ and 
$\sigma_j^\dagger = \frac{1}{2} (X_j - \j Y_j)$
are the spin lowering and raising operators at site $j$, respectively. 
As a result, the Hubbard Hamiltonian is mapped to a sum of Pauli strings,
$\mc {H} = \sum_\ell h_\ell P_\ell$,
where $P_\ell$ are tensor products of Pauli operators acting on qubits.
This Pauli decomposition forms the basis for both Hamiltonian simulation 
and the measurement of Krylov matrix elements discussed in the following 
subsections.

The Hamiltonian respects $\mathrm{U}(1)\times \mathrm{U}(1)$ symmetry, 
leading to the conservation of the total particle number $N_\mr{e}$ and 
the total spin projection $S_z$, and enabling restriction to a fixed 
symmetry sector, such as half-filling.
Exploiting these symmetries substantially reduces the effective Hilbert-space 
dimension, thereby facilitating both classical benchmarking and quantum simulations.
The corresponding quantum circuit for implementing the time-evolution operator
$U(t) = \e^{-\j \mc{H} t}$ is constructed following our previous work \cite{Mirzakhani2025}.

Finally, we note that due to the PBC of our structure, a JW string appears between 
the first and last qubits within each spin sector, as illustrated by the 
rightmost red boxes in Fig.~\ref{fig1}(b). 
This string increases circuit depth and noise, as its implementation typically 
requires multiple CNOT gates \cite{Whitfield2011, Wecker2015solving,Mirzakhani2025}. 
However, as shown in our previous work \cite{Mirzakhani2025}, for a 1D Hubbard 
chain with PBC, this non-local string can be simplified to a global phase 
factor that depends only on the fermionic occupation number within each spin sector. 
This simplification reduces circuit depth and associated noise. 
\hyperref[ap:L_CNOT]{Appendix~\ref{ap:L_CNOT}} quantifies the resulting reduction in 
CNOT gates for different lattice sizes and Trotter numbers.

\subsection{Quantum Krylov diagonalization}  \label{ss:QKD}

Krylov subspace methods are widely used classical techniques for
approximating extremal eigenvalues of large matrices
\cite{Saad2011}. Given a matrix $A$ and a reference vector $|v\rangle$,
the Krylov subspace of order $D$ is defined as
\begin{equation}
    \mc{K}_D(A,|v\rangle) = \mr{span}
    \left\{|v\rangle, A|v\rangle, A^2|v\rangle, \dots, A^{D-1}|v\rangle
    \right\}.
\end{equation}
When $A$ is the Hamiltonian $\mc{H}$ of a quantum system, this construction
generates the so-called \emph{power Krylov subspace}. Classical Krylov
algorithms such as the Lanczos method exploit this subspace to build a
low-dimensional representation of $\mc{H}$ whose extremal eigenvalues
approximate those of the full Hamiltonian.
For many problems, the approximation improves rapidly with increasing Krylov 
dimension, often exhibiting exponential convergence for extremal eigenvalues
\cite{Kaniel1966,Saad1980}.

Because the Hamiltonian itself is generally nonunitary, the classical power 
Krylov construction cannot be implemented directly on a quantum computer.
Instead, QKD constructs a unitary Krylov subspace using the real-time 
evolution operator $U(t)=\e^{-\j \mc{H}t}$, which can be implemented 
using standard Hamiltonian simulation techniques. 
Repeated application of the time-evolution operator
generates a \emph{unitary Krylov subspace}
\begin{equation}
    \mc{K}_U^{D} = \mr{span}
    \left\{|\psi_0 \rangle, U|\psi_0 \rangle, U^2|\psi_0 \rangle, \dots, U^{D-1}|\psi_0 \rangle
    \right\},
\end{equation}
where $|\psi_0 \rangle$ is a chosen reference state and
$U^j = \e^{-\j \mc{H} j t}$ corresponds to time evolution for multiples of a
fixed time step $t$
\cite{Parrish2019,Motta2020,Stair2020,Klymko2022,Cortes2022,Yoshioka2025}. 
Each additional Krylov basis state incorporates information about the system dynamics 
at a later evolution time, thereby systematically enriching the variational 
subspace used to approximate the low-energy eigenspace.
It has been shown that this unitary Krylov
construction retains convergence properties similar to those of
classical power Krylov methods while being compatible with quantum
hardware \cite{Epperly2022}.

QKD exploits this unitary Krylov subspace to obtain approximate 
eigenvalues of the Hamiltonian.
Denoting the Krylov basis states by
\begin{equation}
    |\psi_j\rangle = U^j |\psi_0 \rangle = 
    \e^{-\j \mc{H} j t}| \psi_0 \rangle,
\end{equation}
where $j = 0, 1, \ldots, D-1$, the Hamiltonian projected into the Krylov 
subspace is defined by the matrix elements
\begin{equation} \label{eq:Hmtrx}
    \tilde{\mc H}_{jk} = \langle \psi_j | \mc{H} | \psi_k \rangle
    = \langle \psi_0 | \e^{\j \mc{H} j t} \mc{H} \e^{-\j \mc{H} k t} | \psi_0 \rangle.
\end{equation}
Because the Krylov states are not orthogonal, their overlaps must also be 
measured to construct the overlap (Gram) matrix
\begin{equation} \label{eq:Smtrx}
    \tilde{S}_{jk} = \langle \psi_j | \psi_k \rangle
    = \langle \psi_0 | \e^{\j \mc{H} j t} \e^{-\j \mc{H} k t} | \psi_0 \rangle.
\end{equation}

We note that for the exact time evolutions
\begin{align} \label{}
    \tilde{\mc H}_{jk}
    & = \langle \psi_0 | \mc{H} \e^{-\j \mc{H} (k-j) t} | \psi_0 \rangle
    = \langle \psi_0 | \mc{H} U^{k-j} |\psi_0 \rangle, \label{eq:HmtrxS} \\
    \tilde{S}_{jk} 
    & = \langle \psi_0 | \e^{\j \mc{H} j t} \e^{-\j \mc{H} k t} | \psi_0 \rangle
     = \langle \psi_0 | U^{k-j} |\psi_0 \rangle.  \label{eq:SmtrxS}
\end{align}
Such a simplification leads to a more compact circuit, as it involves only a single time 
evolution \cite{Yoshioka2025}. 
However, when the time evolution is approximated via Trotterization, the two expressions 
are no longer equivalent. 
The expectation values in Eqs.~\eqref{eq:HmtrxS} and \eqref{eq:SmtrxS} can be evaluated 
on a quantum processor using overlap-measurement circuits, such as the Hadamard test 
or related interferometric techniques. 
In this work, we adopt the Hadamard test approach described in Ref.~\cite{Yoshioka2025},
which is reviewed in Sec~\ref{ss:Hada_test}.

The eigenvalue problem for the Hamiltonian within the Krylov subspace
is therefore expressed as the generalized eigenvalue equation
\begin{equation} \label{eq:GeneEig}
    \tilde{\mc H}\mathbf{c} = E\,\tilde{S}\mathbf{c}.
\end{equation}
Diagonalizing the projected Hamiltonian within the Krylov subspace yields approximations 
to the eigenvalues and eigenstates of the full Hamiltonian. 
The lowest eigenvalue provides the QKD estimate of the GSE, while 
the corresponding eigenvector specifies the optimal linear combination of 
Krylov basis states.

In practice, as the Krylov dimension increases, the basis states may become nearly linearly 
dependent, causing the overlap matrix $\tilde{S}$ to become ill-conditioned and 
rendering the generalized eigenvalue problem numerically unstable 
\cite{Stair2020,Klymko2022,Epperly2022}. 
To overcome this difficulty, a regularization procedure based on SVT
is employed \cite{Epperly2022,Kirby2024Ana,Yoshioka2025,Klymko2022,Kirby2023Exact}. 
Specifically, the overlap matrix $\tilde{S}$ is first diagonalized 
(or equivalently decomposed via singular value decomposition), after which both 
$\tilde{H}$ and $\tilde{S}$ are projected onto the subspace spanned by eigenvectors with 
eigenvalues exceeding a prescribed threshold $\delta_{\mathrm{SVT}}>0$. 
Components associated with smaller eigenvalues are discarded, thereby removing nearly 
linearly dependent directions in the Krylov basis and significantly improving the conditioning 
of the projected eigenvalue problem. The resulting truncated generalized eigenvalue problem 
can then be solved in a numerically stable manner.
\begin{figure}
    \centering
    \scalebox{0.95}{
    \begin{quantikz}
        \lstick{$\ket{0}_a$} & \gate{\hat{H}} & \ctrl{1} & \qw & \ctrl[open]{1} & \meter{X \text{ or } Y} \\
        \lstick{$\ket{0}^N$} & \qw           & \gate{\psi_0\text{-prep}} & \gate{U^{k-j}} & \gate{\psi_0\text{-prep}} & \meter{P}
    \end{quantikz}}
    \caption{Hadamard-test circuit used to evaluate the real (via $X$ measurement)
        and imaginary (via $Y$ measurement) components of the Krylov matrix
        element
        $\langle \psi_0 | P U^{k-j} | \psi_0 \rangle$.
        The subcircuit $\psi_0$-prep prepares the reference state
        $|\psi_0\rangle$ from the computational basis, while
        $U^{k-j}$ denotes the time-evolution operator.
        An ancillary qubit, initialized in $|0\rangle$ and acted on by a
        Hadamard gate $\hat{H}$, enables the interference required to extract
        the desired expectation values.
    } \label{fig2}
\end{figure}
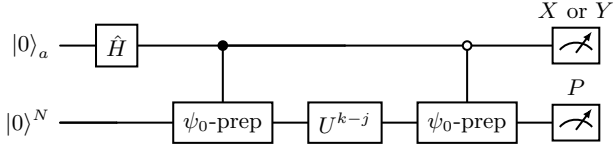

\subsection{Evaluation of Krylov matrix elements} \label{ss:Hada_test}

The implementation of QKD requires the evaluation of the Hamiltonian and overlap matrix 
elements [Eqs.~\eqref{eq:Hmtrx} and \eqref{eq:Smtrx}] within the unitary Krylov subspace. 
These quantities are efficiently computed on a quantum computer using the Hadamard test. 
Throughout this work, we employ the optimized Hadamard-test circuit proposed in 
Ref.~\cite{Yoshioka2025}, illustrated in Fig.~\ref{fig2}.

For an arbitrary Pauli operator $P$, the Hadamard test evaluates the correlation function
\[
\langle \psi_0|P\,U^{k-j}|\psi_0\rangle,
\]
whose real and imaginary parts are obtained from measurements of the auxiliary qubit in 
the $X$ and $Y$ bases, respectively (see \hyperref[ap:Kmatrix]{Appendix~\ref{ap:Kmatrix}}). 
The overlap matrix is recovered by choosing $P=\mathbb{I}$,
\[
\tilde{S}_{jk}
=
\langle\psi_0|U^{k-j}|\psi_0\rangle,
\]
whereas the Hamiltonian matrix is obtained by decomposing the Hamiltonian into Pauli operators,
$\mc{H}=\sum_\ell h_\ell P_\ell$, yielding
\begin{equation}
    \tilde{\mc{H}}_{jk}
    =
    \sum_\ell h_\ell
    \langle\psi_0|P_\ell U^{k-j}|\psi_0\rangle .
\end{equation}

In ideal simulations, the time-evolution operator
$U(t)= \e^{-\j\mc{H} t}$ is implemented exactly. On quantum hardware, however, the Hamiltonian 
is expressed as a sum of generally non-commuting terms,
$\mc{H}=\sum_\ell \mc{H}_\ell$,
and the evolution operator is approximated using the first-order Trotter--Suzuki decomposition~\cite{Hatano2005},
\begin{equation}
    U(t)
    \approx
    \left(
    \prod_\ell
    \e^{-\j \mc{H}_\ell t/N_{\mr{trot}}}
    \right)^{N_{\mr{trot}}},
\end{equation}
where $N_{\mr{trot}}$ denotes the number of Trotter steps. 
Increasing $N_{\mr{trot}}$ systematically reduces the Trotter error at the expense 
of deeper quantum circuits.

Because both $\tilde{\mc{H}}_{jk}$ and $\tilde{S}_{jk}$ depend only on the time 
difference $(k-j)t$, the projected Hamiltonian and overlap matrices possess a Toeplitz structure, 
for example, $\tilde{S}_{jk}=\tilde{S}_{j+1,k+1}$.
Moreover, the unitarity of the time-evolution operator implies
\begin{equation}
    \tilde{S}_{jk}
    =
    \tilde{S}_{kj}^{*},
    \qquad
    \tilde{\mc{H}}_{jk}
    =
    \tilde{\mc{H}}_{kj}^{*},
\end{equation}
so both matrices are Hermitian. 
Exploiting this Toeplitz-Hermitian structure, as in Ref.~\cite{Yoshioka2025}, substantially 
reduces the number of independent matrix elements that must be measured. 
Consequently, the measurement cost scales linearly with the Krylov dimension, 
rather than quadratically, while requiring only a single time-evolution circuit for each 
distinct time difference.

\begin{figure*} [t]
    \centering
    \includegraphics[width=0.65\linewidth]{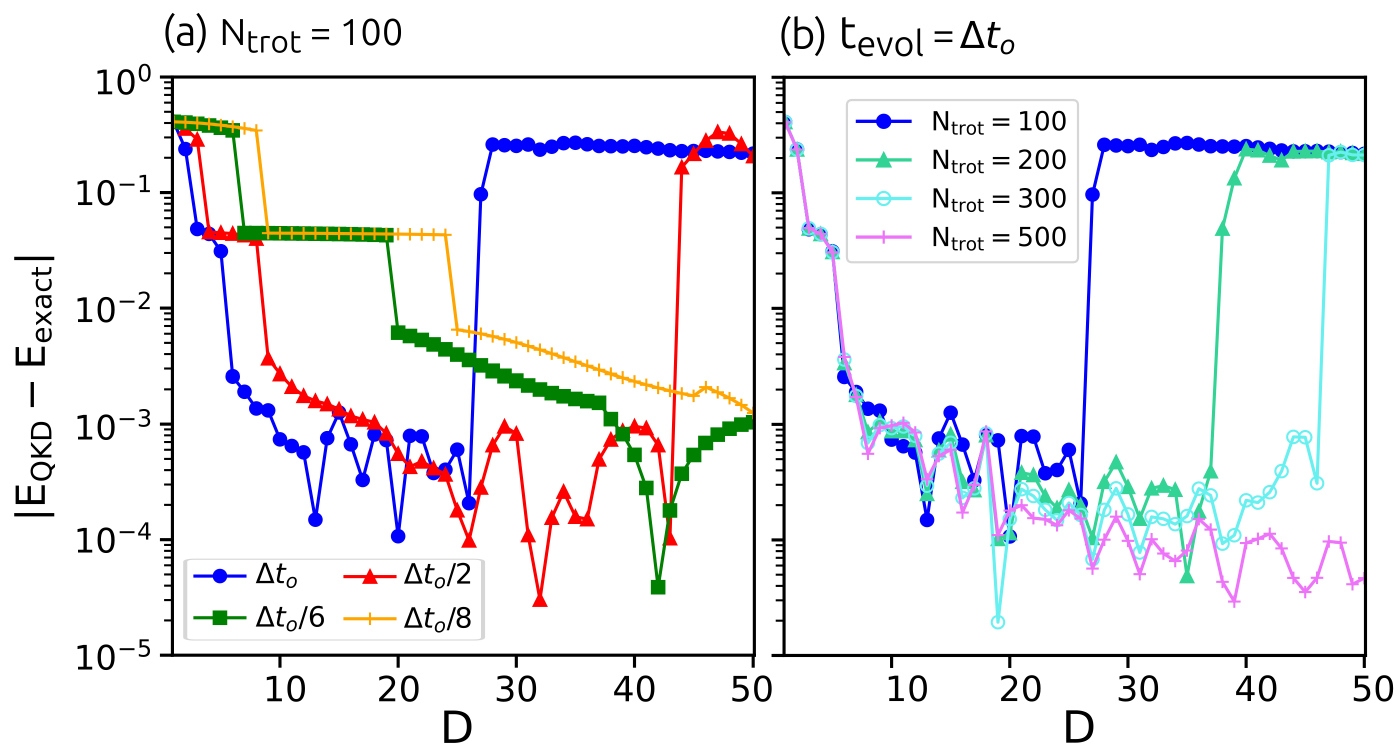}
    \caption{(a) Absolute GSE error as a function of the Krylov dimension
        $D$ for different evolution times $t_\mr{evol}$, as indicated in the
        figure. Results are shown for the weakly interacting case
        $\mc{U}_0=2$ using $N_\mr{trot}=100$ Trotter steps.
        (b) GSE error for the
        $t_\mr{evol}=\Delta t_o$ case shown in panel (a), obtained with
        different Trotter numbers,
        $N_\mr{trot}=100$, $200$, $300$, and $500$.
        The reference evolution time is
        $\Delta t_o=\pi/\|\mc{H}\|\approx0.28$.
    }  \label{fig3}
\end{figure*}

\section{Numerical results} \label{sec:results}

This section presents a systematic investigation of the QKD algorithm through both numerical 
simulations and quantum-hardware experiments.
Leveraging Qiskit \cite{Qiskit}, an open-source quantum computing framework,
we begin by analyzing its convergence properties under ideal noiseless conditions using 
exact statevector simulations, thereby isolating the intrinsic performance of the method 
from statistical and hardware-induced errors. 
The effects of finite-shot sampling and realistic device noise are then examined through 
ideal shot-based simulations and experiments on IBM quantum hardware.

Throughout this work, we consider the half-filled 1D Hubbard model with 
PBC and set the hopping amplitude to $\gamma_0=1$ as the unit 
of energy while varying the interaction strength $\mc{U}_0$. 
For each system size, the exact GSE is obtained using the QuSpin 
package~\cite{Quspin2019} and serves as the benchmark for assessing the QKD results. 
Unless otherwise stated, the reference state $|\psi_0\rangle$ is chosen as the Slater 
determinant corresponding to the noninteracting ground state ($\mc{U}_0=0$), 
the time-evolution operator is implemented using the first-order Trotter--Suzuki 
decomposition with the specified number of Trotter steps, and the singular-value 
truncation threshold is fixed at $\delta_{\mr{SVT}}=0.1$. 
The accuracy of the QKD algorithm is quantified by the absolute GSE 
error, $\Delta E = |E_{\mr{QKD}}-E_{\mr{exact}}|$.

\subsection{Noiseless simulation} \label{ss:NoiseLess}

\subsubsection{Effect of the time step on Krylov convergence} \label{sss:t-eff}

The performance of the QKD algorithm depends critically on the choice of the
time-evolution parameter $t_\mr{evol}$, which determines the structure of the
Krylov subspace. Figure~\ref{fig3}(a) shows the convergence of the QKD
GSE for the half-filled six-site Hubbard model as a function
of the Krylov dimension $D$ for several choices of $t_\mr{evol}$. The evolution
time is expressed relative to the reference scale
$\Delta t_o = \pi / \|\mc H\|$ \cite{Epperly2022,Kirby2024Ana}, where
$\|\mc H\|$ denotes the spectral norm of the Hamiltonian. 
We present results for the weakly interacting case $\mc{U}_0 = 2$ with
$N_{\mr{trot}} = 100$, while the effects of varying interaction strength are 
discussed in Sec.~\ref{sss:U-eff}.

Figure~\ref{fig3}(a) reveals two distinct convergence regimes governed by the
competition between Krylov approximation and time-discretization (Trotter)
errors. 
For sufficiently small evolution times, namely
$t_\mr{evol} = \Delta t_o/6$ and $t_\mr{evol} = \Delta t_o/8$, the energy error
decreases smoothly and monotonically with increasing Krylov dimension. 
In this regime, the Trotter error
is strongly suppressed, and the remaining error is primarily limited by the
finite Krylov subspace. 
Consequently, increasing the Krylov dimension systematically improves 
the ground-state approximation until the finite Krylov-space error 
reaches the Trotter-error floor.

In contrast, for larger evolution times,
$t_\mr{evol} = \Delta t_o$ and $t_\mr{evol} = \Delta t_o/2$, the convergence
behavior changes qualitatively. As observed in Fig.~\ref{fig3}(a), the
plateau-like behavior gradually disappears as $t_\mr{evol}$ increases, while
the energy error initially decreases much more rapidly at small Krylov
dimensions. 
Larger evolution times allow each Krylov basis state to
explore a broader region of Hilbert space, thereby accelerating the initial
convergence. However, they also introduce larger Trotter discretization
errors, which eventually become the dominant source of error. As a result, the
energy error saturates and subsequently exhibits noticeable fluctuations,
eventually increasing with $D$ for the blue and red curves. This behavior
indicates that the accuracy is no longer limited by the Krylov approximation
itself, but by the accumulated errors in the Trotterized time evolution.

To further investigate the breakdown observed at large Krylov dimensions
[Fig.~\ref{fig3}(a)], we fix $t_\mr{evol}=\Delta t_o$ and vary the number
of Trotter steps $N_\mr{trot}$, as shown in Fig.~\ref{fig3}(b).
Increasing $N_\mr{trot}$ systematically delays the onset of the instability:
for $N_\mr{trot}=100$, the error rises sharply beyond $D\!\sim\!25$,
whereas for $N_\mr{trot}=200$ and $300$ the breakdown occurs only at
larger $D$. For $N_\mr{trot}=500$, stable convergence is recovered over
the entire range considered. 
These results indicate that the observed
breakdown is primarily driven by accumulated Trotter discretization
errors. 
Its abrupt onset at large Krylov dimensions is consistent with the increasing 
sensitivity of the generalized eigenvalue problem as the smallest singular 
values of the overlap matrix approach the truncation threshold.
The influence of SVT on the convergence behavior is examined later in this section.

Consistent with the general principle that consecutive Krylov basis states
should contribute new information to the subspace \cite{Klymko2022}, our
results show that choosing a sufficiently large time-evolution parameter
accelerates the initial convergence by generating more informative Krylov
basis states. However, when the time-evolution operator is implemented
through Trotterization, increasing $t_\mr{evol}$ beyond an optimal range
becomes counterproductive due to the accumulation of time-discretization
errors.

These results highlight the trade-off among the evolution time
$t_\mr{evol}$, Krylov dimension $D$, and Trotter number
$N_\mr{trot}$. Smaller evolution times provide more robust convergence
but typically require larger Krylov subspaces, whereas larger evolution
times reduce the required Krylov dimension at the expense of increased
Trotter errors and reduced numerical stability at large $D$.
Consequently, these parameters must be chosen in a balanced manner to
achieve high accuracy while keeping the quantum resources manageable.
For practical NISQ implementations, the optimal choice of $t_\mr{evol}$
is therefore determined not by the fastest apparent convergence, but by 
the best compromise between Krylov efficiency, Trotter accuracy, and circuit depth.

\begin{table}[t]
    \centering
    \caption{Simulation parameters for different lattice sizes $L$ shown in 
    Fig.~\ref{fig4}(a). 
    The time-evolution parameter $t_{\mr{evol}}$ is expressed in units 
    of the reference time scale $\Delta t_o = \pi / \|\mc H\|$, which is defined 
    separately for each system size.}
    \begin{tabular}{c| c c c c c c c c c c}
        \hline\hline
        $L$ & 2 & 3 & 4 & 5 & 6 & 7 & 8 & 9 & 10 \\
        \hline
        $\Delta t_o$ & 0.97 & 0.70 & 0.46 & 0.37 & 0.28 & 0.25 & 0.22 & 0.19 & 0.17 \\
        \hline
        $t_{\mr{evol}}$ 
        & $\frac{1}{6}\Delta t_o$ 
        & $\frac{1}{4}\Delta t_o$ 
        & $\frac{1}{4}\Delta t_o$ 
        & $\Delta t_o$ 
        & $\frac{1}{2}\Delta t_o$ 
        & $\frac{3}{2}\Delta t_o$ 
        & $\frac{1}{2}\Delta t_o$ 
        & $2\Delta t_o$ 
        & $2\Delta t_o$ \\
        & 0.16 & 0.17 & 0.12 & 0.37 & 0.14 & 0.38 & 0.11 & 0.39 & 0.34 \\
        \hline
        $N_{\mr{trot}}$ 
        & 100 & 100 & 200 & 200 & 100 & 200 & 100 & 200 & 200 \\
        \hline\hline
    \end{tabular}
    \label{tab:L-params}
\end{table}
\begin{figure}
    \centering
    \includegraphics[width=0.90\linewidth]{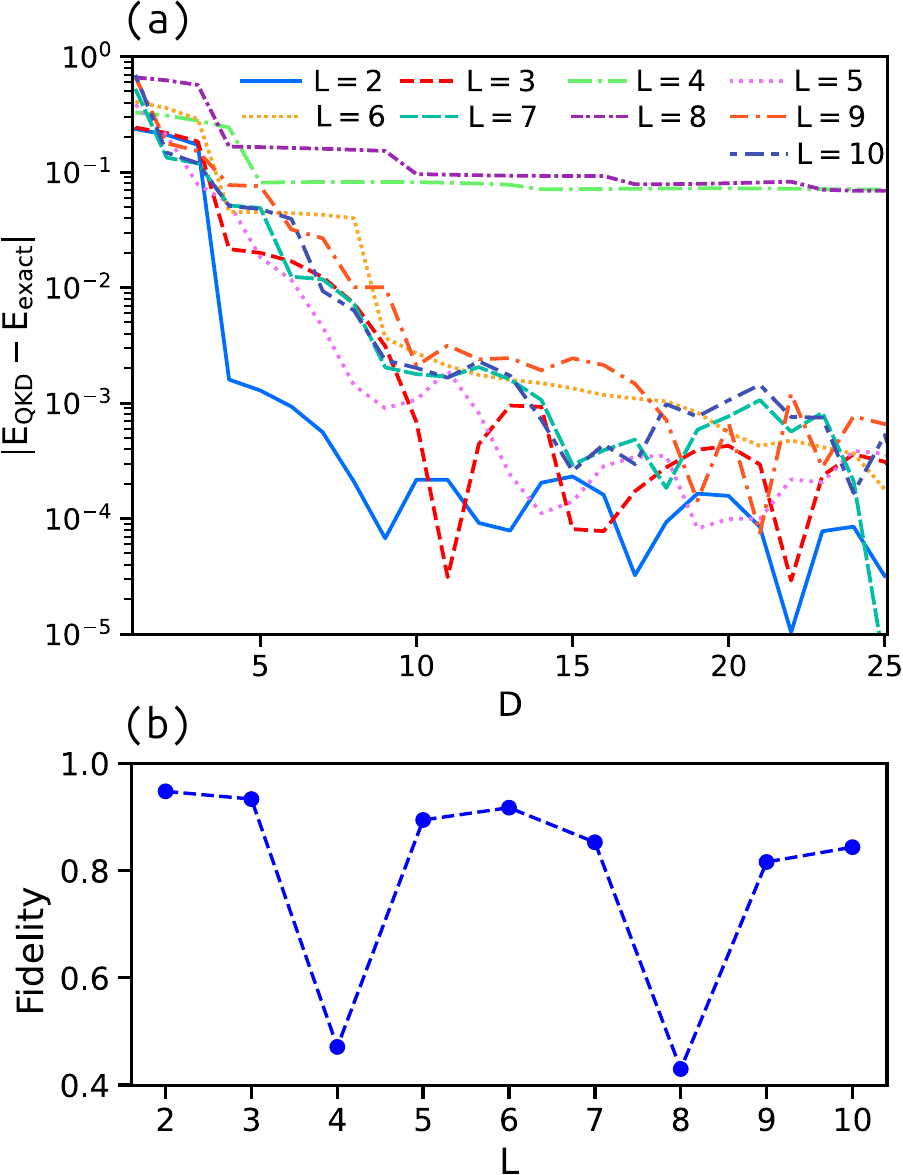}
    \caption{(a) Convergence of the GSE error as a function of the Krylov dimension $D$
        for different lattice sizes $L$.
        For each system size, the time-evolution parameter $t_{\mr{evol}}$ and
        the number of Trotter steps $N_{\mr{trot}}$ are selected relative to the
        reference time scale $\Delta t_o$, as summarized in
        Table~\ref{tab:L-params}, to provide comparable dynamical resolution across
        different lattice sizes.
        (b) Fidelity between the Slater-determinant reference state
        $|\psi_0\rangle$ and the exact ground state $|0\rangle$ of the Hubbard model as a
        function of the lattice size $L$.
        The pronounced reductions in fidelity for $L=4$ and $L=8$ correlate with the
        slower convergence observed in panel (a), indicating a substantially reduced
        overlap between the reference and exact ground states.
    } \label{fig4}
\end{figure}

\subsubsection{System size} \label{sss:Size-eff}
To assess the scalability of the QKD approach, we investigate the
dependence of the algorithmic performance on the system size $L$.
We consider chains of increasing length, $L =2, 3, \ldots  10$, at half-filling,
corresponding to $2L$ qubits in the qubit representation. 
For each system size $L$, the time-evolution parameter $t_{\mr{evol}}$ 
and the number of Trotter steps $N_{\mr{trot}}$ are chosen relative 
to the reference time scale $\Delta t_o = \pi / \|\mc H\|$, which depends 
on the norm of the Hamiltonian for that system. 
This normalization ensures that the time evolution is expressed in units 
that are comparable across different lattice sizes. 
The specific values of $t_{\mr{evol}}$ and $N_{\mr{trot}}$ are then 
adjusted to maintain a similar effective resolution of the dynamics, 
balancing Trotterization error and Krylov convergence. 
In particular, smaller time steps are employed in regimes where larger evolution 
times would otherwise lead to numerical instabilities, whereas larger values 
of $t_\mr{evol}$ are chosen when faster convergence can be achieved without 
significant loss of accuracy.
This choice allows for a consistent comparison across system sizes and ensures 
that meaningful convergence behavior is observed within the accessible Krylov 
dimension range $D \in [1,25]$.
The simulation parameters for each system size are summarized in 
Table~\ref{tab:L-params}.
As in the previous section, we focus on the weakly interacting regime 
with $\mc{U}_0 = 2$.

Figure~\ref{fig4}(a) shows the dependence of the QKD energy error on the Krylov
dimension $D$ for different lattice sizes $L$, revealing a clear system-size dependence
of the convergence.
For relatively small systems ($L=2,3,5$), the error decreases rapidly with increasing
$D$, indicating that a modest Krylov subspace already captures the dominant
low-energy physics.
As the system size increases ($L=6,7,9,10$), the convergence becomes progressively
slower, reflecting the increased complexity of the many-body spectrum.
Consequently, achieving comparable accuracy requires either a larger Krylov
dimension, a more accurate time evolution, or both.

Interestingly, the cases $L=4$ and $L=8$ deviate markedly from the general trend.
Despite using similar simulation parameters, their errors remain substantially larger
and improve only weakly with increasing Krylov dimension.
This behavior is consistent with the general perspective of real-time Krylov methods,
whose efficiency depends strongly on the spectral structure of the Hamiltonian
\cite{Parrish2019,Klymko2022}.
For these two lattice sizes, the relevant low-energy spectrum exhibits
small spectral gaps (see \hyperref[ap:gap]{Appendix~\ref{ap:gap}}), 
reducing the rate at which dynamical phases accumulate.
Consequently, longer evolution times are required to distinguish nearby
eigenstates, so the Krylov basis acquires information about the low-energy
subspace more gradually, leading to slower convergence.
As shown in Fig.~\ref{figA3} of \hyperref[ap:gap]{Appendix~\ref{ap:gap}}, 
employing a longer evolution time indeed improves the convergence for these 
small-gap systems, provided that the time-evolution operator is implemented with 
sufficient accuracy.

Further insight into this behavior is provided by the overlap between the reference
state $|\psi_0\rangle$ and the exact ground state $|0\rangle$.
In Krylov-based approaches, the initial state can be expanded in the eigenbasis  
of the Hamiltonian as
\begin{equation}
    |\psi_0\rangle = \sum_N \gamma_N |N\rangle, 
    \qquad \gamma_N = \langle N | \psi_0 \rangle,
\end{equation}
where $|N \rangle$ are the eigenstates of the Hamiltonian. 
Shown in Fig.~\ref{fig4}(b) is the fidelity between the Slater determinant reference 
state $|\psi_0\rangle$ and the exact ground state of the Hubbard model,
defined as $F = |\langle 0 | \psi_0 \rangle|^2 \equiv |\gamma_0|^2$,
for different lattice sites $L$.
As seen, for most system sizes, this overlap remains sufficiently large ($\gtrsim 0.8$), 
allowing for rapid and stable convergence with increasing Krylov dimension $D$. 
In contrast, for $L=4$ and $L=8$, the fidelity exhibits pronounced drops, signaling 
a substantial reduction in the overlap with the true ground state.

The convergence of QKD is ultimately governed by how rapidly the low-energy eigenstates 
evolves as interactions are introduced. 
When the spectrum remains well separated, the noninteracting Slater determinant retains 
a large overlap with the interacting ground state, allowing the Krylov expansion to 
converge rapidly.
In contrast, when the low-energy gap becomes anomalously small, the ground state 
changes more substantially, reducing the initial overlap and slowing the convergence 
of the Krylov basis.

\begin{figure}
    \centering
    \includegraphics[width=0.95\linewidth]{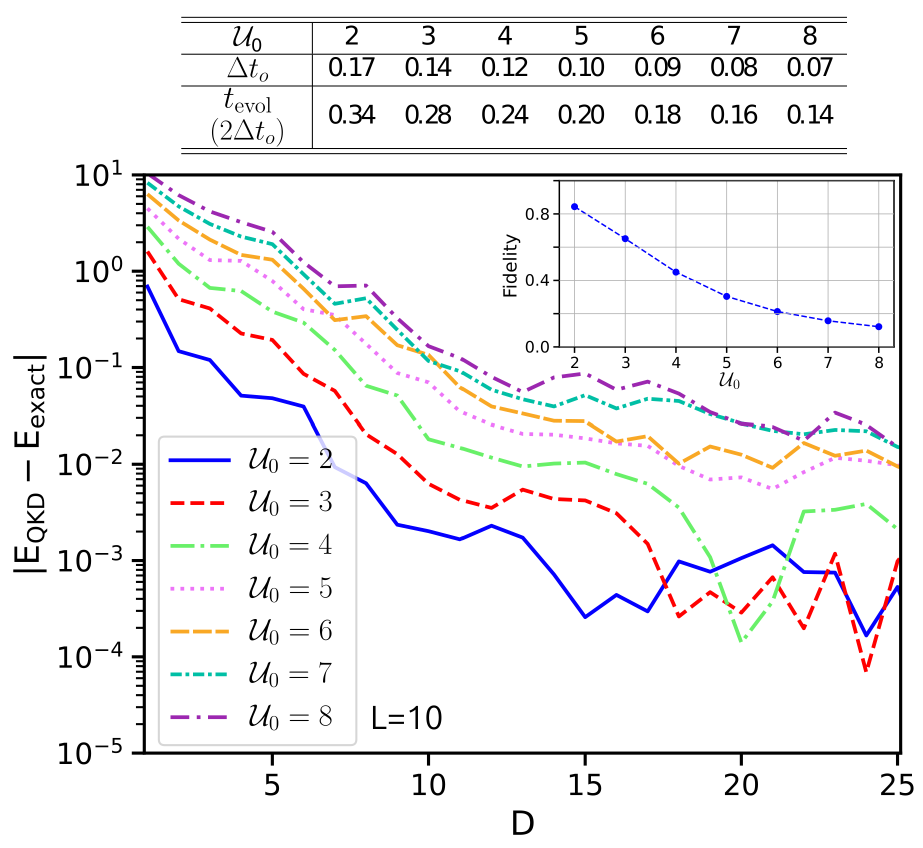}
    \caption{Absolute GSE error as a function of the Krylov dimension $D$ 
        for different on-site Coulomb interaction strengths
        $\mc{U}_0=2$--$8$ at fixed system size $L=10$.
        For each value of $\mc{U}_0$, the time-evolution parameter is chosen as
        $t_\mr{evol}=2\Delta t_o$ (see the table at the top of the figure), with
        $N_\mr{trot}=200$ Trotter steps.
        The inset shows the fidelity between the Slater-determinant reference state
        and the true ground state as a function of $\mc{U}_0$, illustrating the reduction in
        their overlap as the interaction strength increases.
    } \label{fig5}
\end{figure}

\subsubsection{Coulomb interaction} \label{sss:U-eff}

We next investigate the influence of the on-site Coulomb interaction
strength on the convergence of QKD by varying $\mc{U}_0$ from $2$ to $8$.
Throughout this analysis, the system size is fixed at $L=10$, while the
time evolution is performed with $t_\mr{evol}=2\Delta t_o$ using
$N_\mr{trot}=200$ Trotter steps.
This allows us to isolate the effect of increasing electronic
correlations on the performance of the algorithm.

Figure~\ref{fig5} shows the GSE error as a function of
the Krylov dimension $D$ for different interaction strengths.
For weak interactions ($\mc{U}_0=2,3,4$), the error decreases rapidly
with increasing $D$, demonstrating efficient convergence.
As the interaction strength increases, however, the convergence becomes
progressively slower and eventually saturates at larger errors.

This behavior can be understood from the changing character of the
ground state.
The reference state employed throughout this work is the
noninteracting Slater determinant, which provides an excellent
approximation in the weakly correlated regime.
As $\mc{U}_0$ increases, the exact ground state becomes increasingly
correlated, reducing its overlap with the reference state.
This trend is confirmed by the fidelity shown in the inset of
Fig.~\ref{fig5}, which decreases monotonically with increasing
$\mc{U}_0$.
Consequently, the Krylov basis contains less information about the true
ground state from the outset, requiring larger Krylov subspaces to
recover the missing correlations and leading to slower convergence.
\begin{figure}
    \centering
    \includegraphics[width=0.9\linewidth]{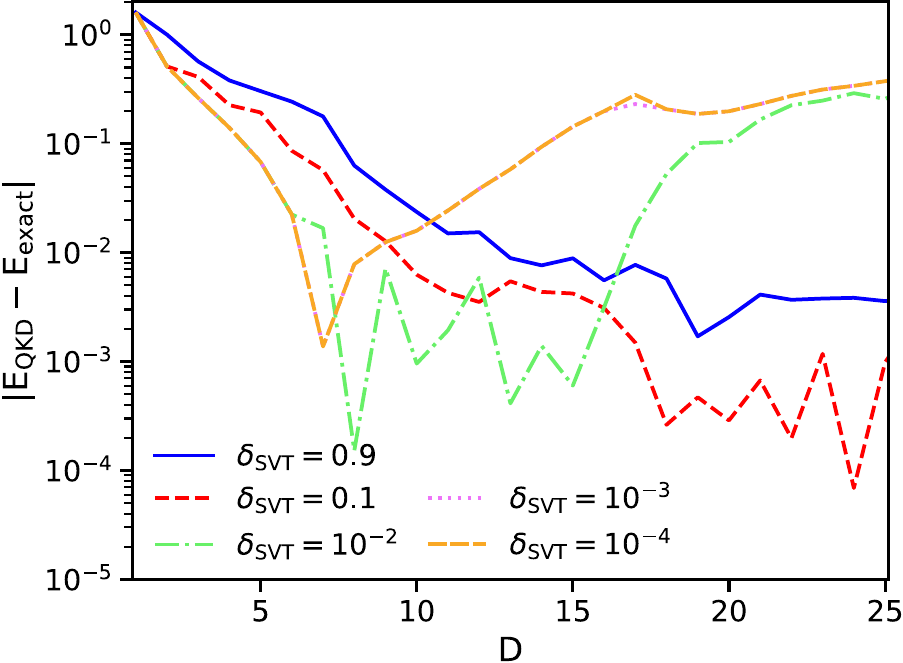}
    \caption{Absolute GSE error as a function of the Krylov dimension $D$ for different
        SVT thresholds $\delta_{\mr{SVT}}$ at $L=10$ and $\mc{U}_0=3$.
        The time evolution is performed with
        $t_\mr{evol}=2\Delta t_o=0.28$ using
        $N_\mr{trot}=200$ Trotter steps.
        Intermediate truncation thresholds (red) yield the lowest errors, whereas
        larger thresholds (blue) degrade the convergence by discarding relevant Krylov
        basis components, and smaller thresholds lead to numerical instabilities
        at larger Krylov dimensions.
    } \label{fig6}
\end{figure}

\subsubsection{Singular-value truncation} \label{sss:SVT}

We finally investigate the role of the SVT threshold, which regularizes 
the generalized eigenvalue problem by discarding Krylov basis vectors 
associated with small singular values of the overlap matrix.
As discussed in Sec.~\ref{ss:QKD}, this regularization improves the
conditioning of the overlap matrix but may also remove physically
relevant information if the truncation is too aggressive.
The optimal choice of $\delta_{\mr{SVT}}$ therefore reflects a balance
between numerical stability and the completeness of the Krylov basis.

To examine this trade-off, we consider a system with $L=10$ and
$\mc{U}_0=3$, using $t_{\mr{evol}}=2\Delta t_o$ and
$N_{\mr{trot}}=200$.
The resulting convergence as a function of the Krylov dimension is shown
in Fig.~\ref{fig6}.

For large truncation thresholds, such as
$\delta_{\mr{SVT}}=0.9$, the energy error decreases only modestly with
increasing Krylov dimension.
In this regime, many Krylov basis vectors are discarded, reducing the
effective dimension of the subspace and preventing the variational space
from systematically improving.
Reducing the threshold to $\delta_{\mr{SVT}}=0.1$ enhances
the convergence, yielding the smallest errors over the entire range of
Krylov dimensions considered.

A different behavior emerges for very small truncation thresholds
($\delta_{\mr{SVT}}=10^{-2},\,10^{-3},\,10^{-4}$).
Although the initial convergence is faster because more Krylov vectors
are retained, the error increases again at larger Krylov dimensions.
This deterioration arises because nearly linearly dependent basis states
are no longer removed, causing the overlap matrix to become
progressively ill-conditioned and amplifying numerical errors in the
generalized eigenvalue problem.
This behavior is consistent with the instability discussed previously
for large Krylov dimensions, where small singular values make the QKD
solution increasingly sensitive to errors.

Overall, these results demonstrate that neither excessive nor
insufficient truncation is desirable.
An intermediate truncation threshold provides the best compromise,
retaining the physically relevant Krylov subspace while suppressing the
small-singular-value modes responsible for numerical instability.

\begin{table}[h]
    \centering
    \caption{Comparison of circuit resources for QKD and IQPE applied to the six-site
    Hubbard model ($L=6$, $\mc{U}_0=3$). 
    For IQPE, the execution parameters are $m=5$ binary digits and $N_{\mr{trot}}=10$ 
    Trotter steps, where circuit depth increases with the iteration index $k$. 
    For QKD, the results are obtained with Krylov subspace dimension $D=15$, 
    $N_{\mr{trot}}=50$, and a time step $t_{\mr{evol}} = \Delta t_0 \approx 0.23$.
    Notably, the QKD circuit structure remains fixed across all Krylov dimensions, 
    with accuracy improved through classical post-processing rather than 
    increased circuit depth.}
    \begin{tabular}{c|c|c|c|c}
    \hline \hline
    \textbf{Method} & $\mathbf{k}$ & \textbf{Circuit size} & \textbf{Depth} & \textbf{Width} \\
    \hline
         & 4 & 279110 & 205196 & 14 \\
         & 3 & 139830 & 102636 & 14 \\
    IQPE & 2 & 70190  & 51356  & 14 \\
         & 1 & 35370  & 25716  & 14 \\
         & 0 & 17960  & 12896  & 14 \\
    \hline
    QKD & (each $D$) & 15435 & 5943 & 13 \\
    \hline \hline
    \end{tabular}
    \label{tab:iqpe_qkd_com}
\end{table}

\subsubsection{Comparison with Iterative Quantum Phase Estimation} \label{ss:QKD_iqpe}

To further assess the practical performance of QKD, we compare it with
IQPE \cite{Mirzakhani2025} for the
$L=6$ Hubbard model at $\mc{U}_0=3$ under noiseless conditions.
The simulation parameters for both methods are summarized in
Table~\ref{tab:iqpe_qkd_com}.

\begin{figure}
    \centering
    \includegraphics[width=0.8\linewidth]{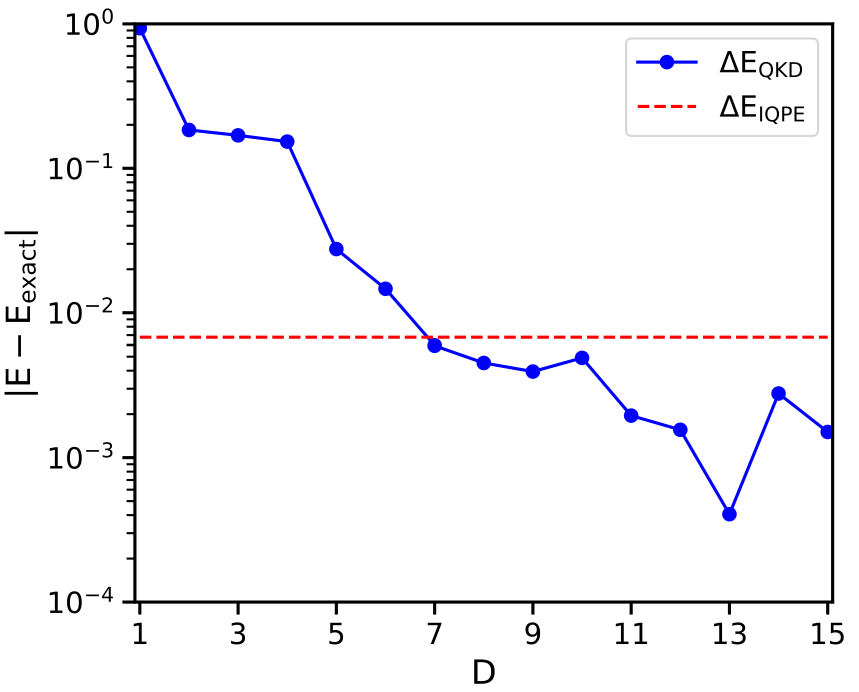}
    \caption{Comparison of the absolute GSE error obtained with QKD (blue circles) and
        IQPE (red dashed line) for the six-site Hubbard model
        ($L=6$, $\mc{U}_0=3$, $\delta_\mr{SVT}=0.1$).
        The IQPE result corresponds to the simulation parameters listed in
        Table~\ref{tab:iqpe_qkd_com}.
        QKD achieves higher accuracy than IQPE for $D\gtrsim7$ while employing
        substantially shallower quantum circuits.
    } \label{fig7}
\end{figure}

Figure~\ref{fig7} compares the absolute GSE errors of the
two approaches.
IQPE yields an error of approximately
$7\times10^{-3}$ with the chosen simulation parameters.
In contrast, QKD rapidly improves as the Krylov dimension increases and
already surpasses the IQPE accuracy at $D\approx7$, ultimately reaching
errors of $\mathcal{O}(10^{-3}\text{--}10^{-4})$ for
$D\approx7$--15.

The improved accuracy is accompanied by a substantially lower circuit
complexity.
As shown in Table~\ref{tab:iqpe_qkd_com}, IQPE requires deep quantum
circuits, with depths exceeding $2\times10^5$ for the most significant
controlled time evolutions, owing to the iterative phase-accumulation
procedure.
By contrast, the depth of the Hadamard-test circuits used in QKD remains
approximately $6\times10^3$ and is essentially independent of the Krylov
dimension.
Improving the QKD accuracy therefore relies primarily on enlarging the
classically processed Krylov subspace rather than increasing the quantum
circuit depth.

This comparison highlights a fundamental distinction between the two
approaches.
IQPE improves precision by performing increasingly long coherent quantum
evolutions, making it highly sensitive to circuit errors and coherence
limitations.
QKD instead transfers much of the computational effort to classical
post-processing while keeping the quantum circuits relatively shallow.
For the problem considered here, this strategy provides both higher
accuracy and significantly lower quantum resource requirements, making
QKD a particularly attractive eigensolver for current NISQ quantum
processors.

\begin{figure*}
    \centering
    \includegraphics[width=0.8\linewidth]{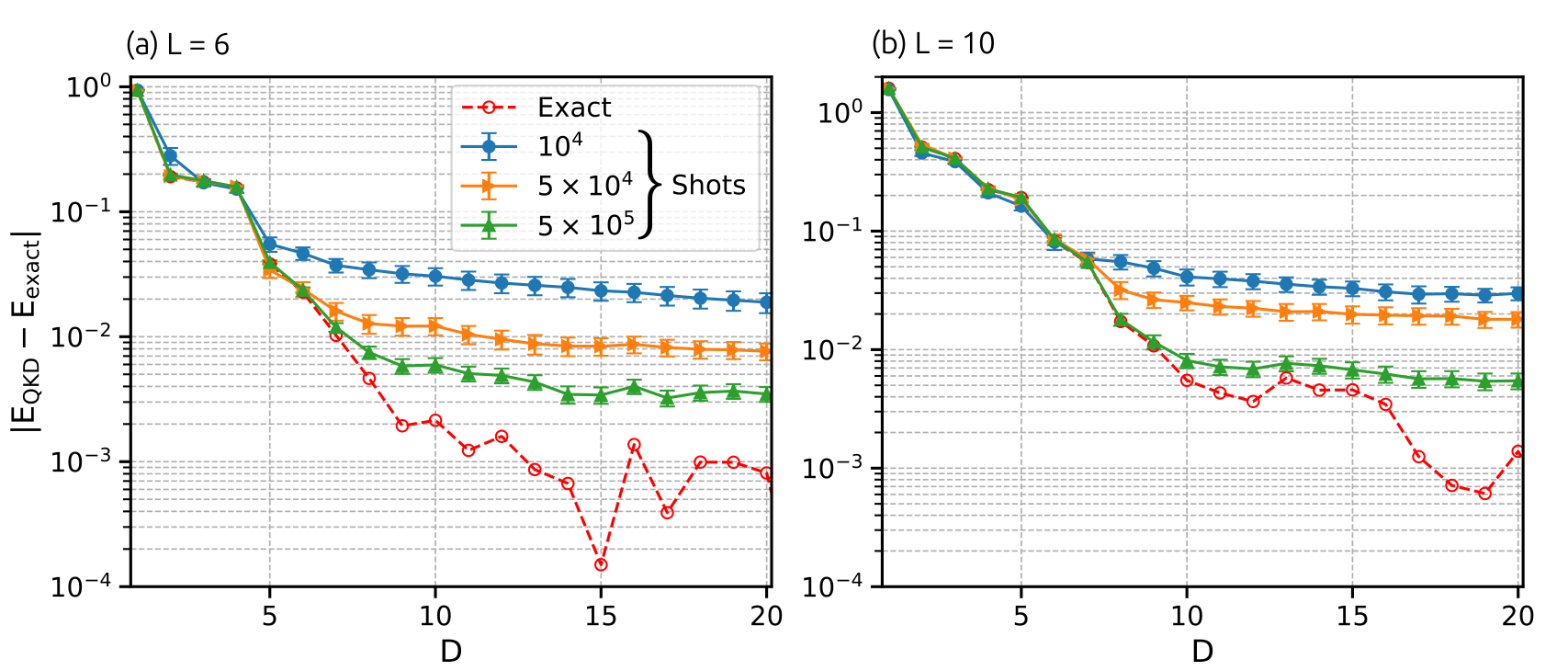}
    \caption{Absolute GSE error as a function of the Krylov dimension $D$ for the
        Hubbard model with (a) $L=6$ and (b) $L=10$ lattice sites at
        $\mc{U}_0=3$.
        The red dashed curves denote exact simulations, whereas the blue,
        orange, and green curves correspond to shot-based QKD simulations
        using $10^4$, $5\times10^4$, and $5\times10^5$ measurement shots,
        respectively.
        The time evolution is performed with
        $t_\mr{evol}=\Delta t_o=0.23$ for $L=6$ and
        $t_\mr{evol}=2\Delta t_o=0.28$ for $L=10$, using
        $N_\mr{trot}=100$ and $150$ Trotter steps, respectively.
        The SVT threshold is fixed at
        $\delta_\mr{SVT}=0.1$.
        Error bars represent one bootstrap standard deviation estimated from
        25 independent realizations 
        (see \hyperref[ap:bootstrap]{Appendix~\ref{ap:bootstrap}}).
    } \label{fig8}
\end{figure*}

\subsection{Noisy simulation and experiment} \label{ss:shot_noise}

\subsubsection{Finite-shot sampling effects} \label{sss:shot}

The noiseless simulations presented in the previous section isolate the intrinsic
convergence properties of the QKD algorithm. In practical quantum processors,
however, expectation values are estimated from a finite number of circuit
executions, introducing statistical fluctuations in the measured Hamiltonian and
overlap matrix elements. These fluctuations propagate through the generalized
eigenvalue problem and ultimately limit the achievable energy accuracy.
Understanding this sampling-induced error is therefore an important step toward
assessing the practical performance of QKD before considering realistic hardware
noise.

To isolate the effect of finite-shot statistics, we perform shot-based QKD
simulations in the absence of gate and decoherence errors for Hubbard chains of
sizes $L=6$ and $L=10$ at $\mc{U}_0=3$, as shown in Fig.~\ref{fig8}. Several
measurement budgets are considered in order to examine the statistical
convergence of the algorithm. Statistical uncertainties are estimated using
bootstrap resampling (\hyperref[ap:bootstrap]{Appendix~\ref{ap:bootstrap}}) 
over independent sampling realizations. 
The detailed simulation parameters and shot configurations are given in the 
caption of Fig.~\ref{fig8}.

For both system sizes, the energy error decreases rapidly with increasing Krylov
dimension in the low-$D$ regime ($D\lesssim7$), closely following the exact
statevector results. This demonstrates that the early-stage convergence remains
governed by the systematic improvement of the Krylov approximation, while
sampling noise has only a minor influence on the estimated matrix elements.
Consequently, the conclusions drawn from the noiseless analysis remain valid in
this regime.

As the Krylov dimension increases, the behavior changes qualitatively.
Whereas the exact simulations continue to improve, eventually reaching
sub-$10^{-3}$ accuracy, the shot-based calculations gradually saturate and no
longer benefit significantly from enlarging the Krylov subspace. In this regime,
the dominant source of error is no longer the finite Krylov approximation, but
the statistical uncertainty of the measured Hamiltonian and overlap matrices.
Increasing the number of measurement shots systematically lowers this accuracy
floor, consistent with the expected $1/\sqrt{N_\mr{shots}}$ scaling of sampling
noise.

A comparison between Figs.~\ref{fig8}(a) and \ref{fig8}(b) further shows that
this behavior is qualitatively unchanged as the system size increases from
$L=6$ to $L=10$. Although the larger system exhibits slightly larger residual
errors at high Krylov dimensions, no significant deterioration of the
convergence mechanism is observed. Instead, both systems display the same
transition from a Krylov-limited regime at small $D$ to a sampling-limited
regime at larger $D$. This observation indicates that finite-shot statistics,
rather than the Krylov approximation itself, become the dominant factor
determining the achievable accuracy once the Krylov subspace is sufficiently
large.

Overall, these results identify a practical operating regime for QKD on
near-term quantum devices. Increasing the Krylov dimension improves the energy
estimate only until the statistical uncertainty of the measured matrix elements
becomes comparable to the remaining systematic error. Beyond this point,
additional Krylov basis states provide diminishing returns unless accompanied by
a corresponding increase in the measurement budget.

\begin{figure*}
    \centering
    \includegraphics[width=0.8\linewidth]{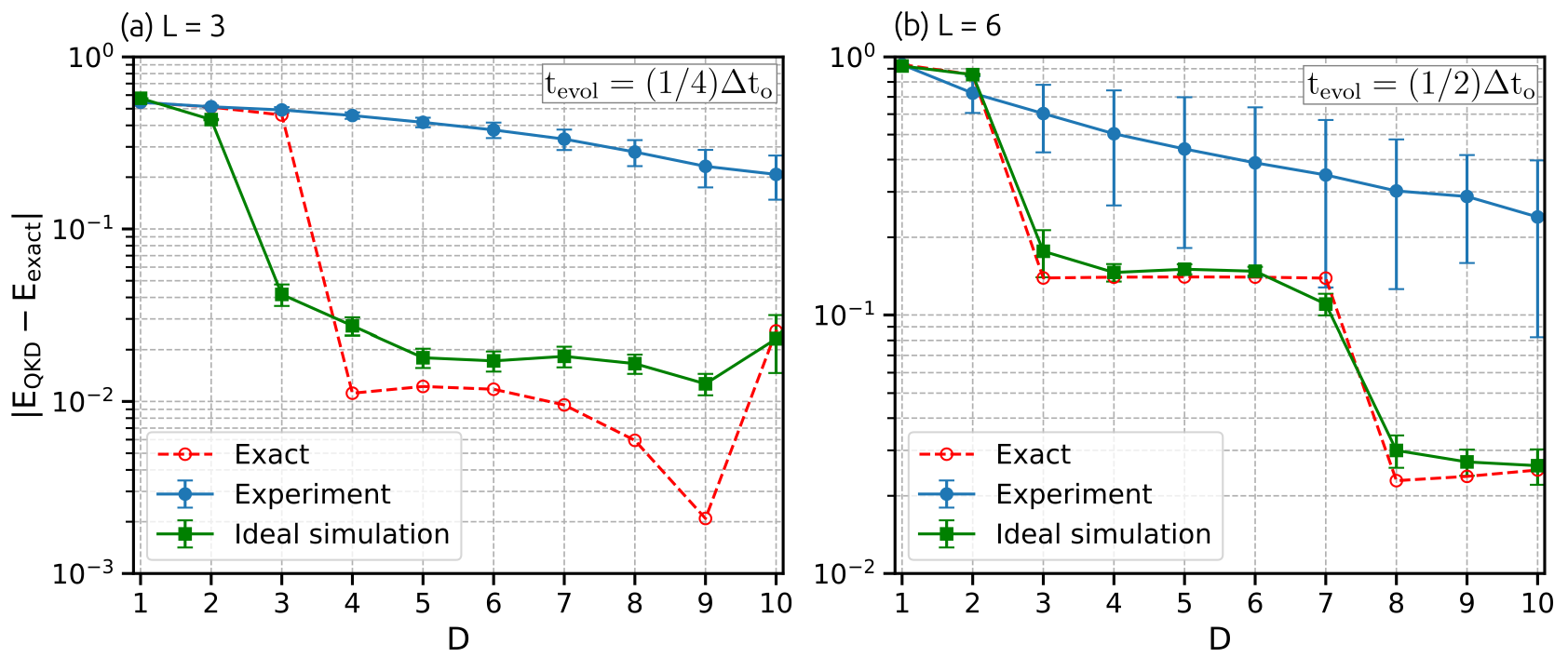}
    \caption{Absolute GSE error as a function of the Krylov dimension $D$ for the
        Hubbard model with (a) $L=3$ and (b) $L=6$ at $\mc{U}_0=3$.
        Results obtained from exact calculations (red open circles), ideal
        shot-based simulations (green squares), and experiments performed on
        the \texttt{ibm\_yonsei} quantum processor (blue filled circles) are
        compared.
        The time-evolution parameter is chosen as
        $t_\mr{evol}=\Delta t_o/4\approx0.15$ for $L=3$ and
        $t_\mr{evol}=\Delta t_o/2\approx0.12$ for $L=6$.
        Experimental error bars represent one standard deviation estimated from
        six independent hardware runs, each using $10^4$ measurement shots.
        Error bars for the ideal shot-based simulations represent one bootstrap
        standard deviation estimated from 25 independent realizations, each
        using $10^4$ measurement shots.
    } \label{fig9}
\end{figure*}

\subsubsection{Realistic device-noise simulations} \label{sss:device_sim}

Finally, we assess the performance of the QKD algorithm under realistic
hardware conditions using the IBM quantum processor \texttt{ibm\_yonsei},
based on IBM's 127-qubit Eagle architecture.
We compute the GSE of the half-filled Hubbard model for
two representative system sizes, $L=3$ and $L=6$, at $\mc{U}_0=3$.
To minimize the experimental overhead, we employ only a lightweight
readout-error mitigation protocol based on Twirled Readout Error
Extinction (TREX)~\cite{Van2022}, which suppresses measurement errors
through randomized measurement twirling followed by the corresponding
classical bit-flip correction.

Figure~\ref{fig9} compares the convergence of the QKD algorithm obtained
from exact calculations, ideal shot-based simulations, and executions on
IBM quantum hardware.
The experimental data correspond to six independent hardware runs,
each using $10^4$ measurement shots.
The error bars represent one standard deviation over these six runs.
For comparison, the ideal shot-based simulations employ the same shot
budget, with statistical uncertainties estimated from 25 independent
sampling realizations using bootstrap resampling.

For both system sizes, the ideal shot-based simulations closely reproduce
the exact results, confirming the conclusions of the previous subsection
that finite-shot statistics alone have only a limited impact on the
convergence.
Although the hardware results exhibit larger errors due to device noise,
they preserve the same overall convergence behavior, with the
GSE improving systematically as the Krylov dimension
increases.
The experimental data therefore indicate that the convergence mechanism
identified in the noiseless and shot-based analyses remains observable
even on present-day superconducting quantum hardware.

The remaining discrepancy between the experimental and ideal results
originates primarily from hardware imperfections, including two-qubit
gate errors, decoherence, state-preparation errors, and residual
measurement errors.
These effects accumulate as the Krylov dimension increases because
higher-order Krylov basis states require longer time-evolution circuits.
Consequently, both the energy error and its statistical spread increase
with system size and Krylov dimension, as reflected by the larger error
bars in Fig.~\ref{fig9}.
Nevertheless, even without advanced error-mitigation techniques such as
zero-noise extrapolation (ZNE)~\cite{Temme2017,Li2017} or probabilistic error
cancellation (PEC)~\cite{Temme2017,Endo2018,Takagi2022}, the QKD algorithm reproduces the
expected convergence trend using only $10^4$ shots per circuit.

The present hardware demonstration highlights an important practical
feature of the QKD approach.
Unlike variational algorithms such as VQE, QKD avoids iterative
quantum-classical optimization and instead constructs a predetermined
Krylov basis through real-time evolution.
Consequently, the experimental cost is determined primarily by the
execution of a finite set of quantum circuits, while the final energy is
obtained through classical diagonalization of a relatively small
generalized eigenvalue problem.
Together with the favorable agreement between experiment and simulation,
these results indicate that QKD provides a promising and hardware-efficient
route toward ground-state calculations on current NISQ quantum
processors.

\section{Conclusion}  \label{sec:conc}

In this work, we have presented a comprehensive study of the QKD algorithm 
for computing the GSE of the half-filled 1D Hubbard model on both 
ideal and real quantum devices. 
Building upon our simplified JW implementation, which
reduces the number of CNOT gates required for quantum time evolution, we
have analyzed the algorithm from both numerical and practical perspectives,
thereby improving its suitability for NISQ-era quantum hardware.

Through extensive numerical simulations, we systematically investigated the
influence of the time-evolution parameter, Krylov dimension, Trotter number,
SVT threshold, system size, and interaction strength on the convergence of QKD. 
Our results show that the performance of the method is
governed by the competition between Krylov-space expressibility, numerical
stability, and the accuracy of the implemented time evolution. 
In particular, we showed that the convergence rate is strongly governed by the 
low-energy spectral structure of the Hamiltonian. 
Near-closing energy gaps slow the accumulation of spectral information within 
the Krylov subspace, requiring longer evolution times to resolve closely spaced 
eigenstates. 
At the same time, the evolution time, Krylov dimension, Trotter number, and 
SVT threshold must be carefully balanced to achieve high accuracy while 
maintaining numerical stability.
We further demonstrated that increasing the Trotter number extends the range 
of stable Krylov convergence by suppressing accumulated time-discretization errors.

The practical applicability of the method was demonstrated through both
shot-based simulations and executions on the IBM \texttt{ibm\_yonsei} quantum
processor. 
While finite-shot statistics introduce an accuracy floor and
hardware noise limits the attainable precision, the characteristic convergence
behavior predicted by the noiseless analysis remains clearly observable.
Using only lightweight readout-error mitigation and a modest measurement
budget, the experimental results reproduce the convergence trends obtained from
ideal simulations, demonstrating that meaningful ground-state approximations
can already be achieved on present-day superconducting quantum hardware.

Beyond benchmarking the algorithm, this work provides practical guidelines for
selecting the key parameters governing QKD, clarifying the trade-offs among
convergence rate, numerical stability, circuit depth, and experimental cost.
These insights should facilitate efficient implementations of Krylov-based
quantum eigensolvers for larger and more strongly correlated quantum systems.

Overall, our results establish QKD as a practical and hardware-efficient
approach for quantum many-body simulations on current NISQ processors.
As quantum hardware, Hamiltonian-simulation methods, and error-mitigation
techniques continue to advance, Krylov-based quantum algorithms are expected to
become an increasingly powerful framework for studying strongly correlated
fermionic systems beyond the reach of classical computation.

\section*{Acknowledgment} \label{ack}
HK and MM were supported by the IITP (RS-2025-25464252, RS-2024-00437191) and 
the NRF (RS-2025-25464492, RS-2024-00442710) funded by the Ministry of 
Science and ICT (MSIT), Korea.
KM and MM were supported by the NRF of Korea (NRF-2022M3K2A108385811, RS-2023-00257561).

\appendix

\section{CNOT gate counts for standard and simplified JW mappings} \label{ap:L_CNOT}
In this appendix, we report the number of CNOT gates for the standard and simplified 
JW mappings in a 1D Hubbard chain with PBC, evaluated directly from the constructed 
circuits prior to any transpilation. 
This choice isolates the intrinsic gate overhead associated with each implementation, 
independent of hardware-specific optimizations or compiler-induced transformations. 
In particular, it allows us to explicitly quantify the reduction in 
CNOT gate count arising from the simplification of the non-local 
JW string under PBC, providing a transparent comparison of the underlying 
circuit constructions.

\begin{figure}
    \centering
    \includegraphics[width=0.85\linewidth]{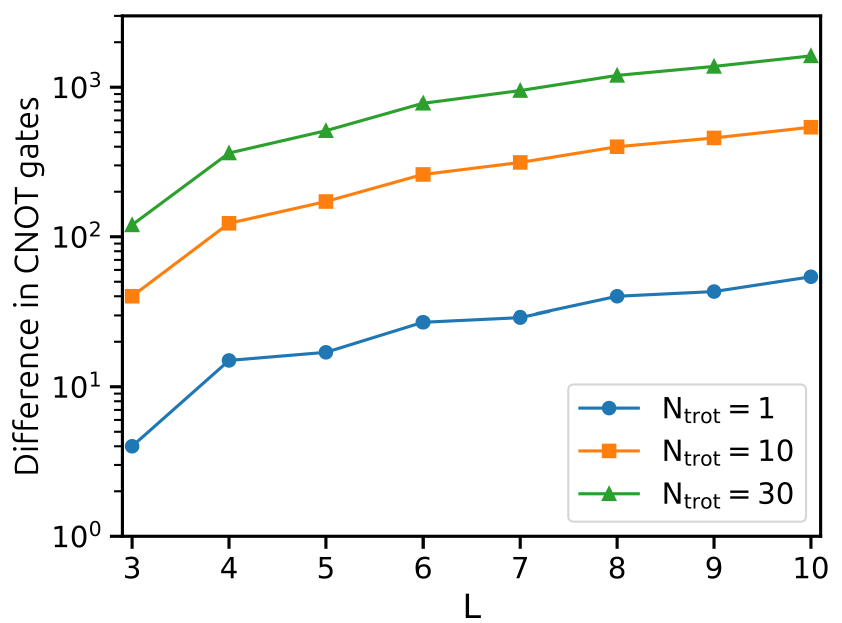}
    \caption{Reduction in the number of CNOT gates as a function of the 
        lattice size $L$ for different Trotter numbers $N_{\mr{trot}} = 1, 10, 30$. 
        The reported values correspond to the difference between the standard and 
        simplified circuit constructions for a single Krylov step; 
        this reduction is identical across all Krylov dimensions.
        } \label{figA1}
\end{figure}

Figure~\ref{figA1} quantifies the reduction in CNOT gates
achieved by the simplified circuit construction relative to the standard
JW implementation.
The number of eliminated CNOT gates increases with both the lattice size
$L$ and the number of Trotter steps $N_{\mr{trot}}$, reflecting the
growing contribution of the nonlocal JW strings to the
time-evolution circuit.
Since this reduction is obtained for every implementation of the
time-evolution operator, the cumulative gate savings increase
approximately linearly with the number of Krylov basis states employed
in the QKD calculation.
Combined with the Toeplitz-based evaluation of the Krylov matrices
described in Sec.~\ref{ss:Hada_test}, this reduction further lowers the overall
computational resources required by the algorithm.
From a practical perspective, reducing the number of two-qubit (CNOT) gates is
particularly important for current superconducting quantum processors,
where such gates typically exhibit the largest error rates and longest
execution times, thereby improving the feasibility of QKD on NISQ
hardware.

\section{Evaluation of Krylov Matrix Elements via Hadamard Test} \label{ap:Kmatrix}

We assume that the computational basis state
$|0\rangle^{\otimes N}$ is an eigenstate of the Hamiltonian $\mc H$. 
This condition is satisfied for Hamiltonians preserving
$U(1)$ symmetry (e.g., particle-number conservation), where
$|0\rangle^{\otimes N}$ represents the vacuum state.
The corresponding eigenvalue can be computed efficiently from the
Pauli decomposition of $\mc H$.
Consequently, the phase accumulated under the time-evolution operator,
including its Trotterized implementation, is known exactly as long as
the symmetry is preserved \cite{Yoshioka2025}.

The Hadamard-test circuit shown in Fig.~\ref{fig2} prepares the
following state immediately before the final measurement:
\begin{align}
    |0 \rangle_a |0\rangle^{\otimes N} &\xrightarrow{\hat{H}} 
    \frac{1}{\sqrt{2}} \left[|0\rangle_a |0\rangle^{\otimes N} + 
    |1\rangle_a |0\rangle^{\otimes N}\right] \notag \\
    &\xrightarrow{\text{1-ctrl-prep}} \frac{1}{\sqrt{2}} 
    \left[|0\rangle_a |0\rangle^{\otimes N} + |1\rangle_a |\psi_0\rangle\right] \notag \\
    &\xrightarrow{U^{k-j}} \frac{1}{\sqrt{2}} \left[\e^{\j \phi} |0\rangle_a |0\rangle^{\otimes N} + |1\rangle_a U^{k-j} |\psi_0\rangle \right] \notag \\
    &\xrightarrow{\text{0-ctrl-prep}} \frac{1}{\sqrt{2}} 
    \left[\e ^{\j \phi} |0\rangle_a |\psi_0\rangle + |1\rangle_a U^{k-j} 
    |\psi_0\rangle \right], 
\end{align}
where the phase shift $U^{k-j} |0\rangle^{\otimes N} = \e^{\j \phi} |0\rangle^{\otimes N}$ 
has been used in the third line.
Expressing the ancilla qubit in the eigenbases of the Pauli-$X$ and
Pauli-$Y$ operators yields
\begin{multline}    
    \frac{1}{2} 
    ( 
    |+\rangle_a \left[\e^{\j \phi} |\psi_0\rangle + U^{k-j} |\psi_0\rangle \right] \\
    + |-\rangle_a \left[\e^{\j\phi} |\psi_0\rangle - U^{k-j} |\psi_0\rangle \right] 
    ), 
\end{multline}
\vspace{-0.7cm}
\begin{multline}
    \frac{1}{2} 
    (
    |+\j \rangle_a \left[\e^{\j\phi} |\psi_0\rangle - \j\, U^{k-j} |\psi_0\rangle \right] \\
    +|-\j \rangle_a \left[\e^{\j\phi} |\psi_0\rangle + \j\, U^{k-j} |\psi_0\rangle \right]
    ).
\end{multline}
%
Therefore, measuring the observables
$X\otimes P$ and $Y\otimes P$
gives
\begin{align}
    \langle X \otimes P \rangle &= \mathrm{Re}\left[ \e^{-\j \phi} \langle \psi_0 | P U^{k-j} | \psi_0 \rangle \right], \\
    \langle Y \otimes P \rangle &= \mathrm{Im}\left[ \e^{-\j \phi} \langle \psi_0 | P U^{k-j} | \psi_0 \rangle \right].
\end{align}
Since the phase $\phi$ is known classically, it can be removed
during post-processing, allowing the desired matrix elements
$\langle\psi_0|PU^{k-j}|\psi_0\rangle$
to be reconstructed directly.

The Hamiltonian matrix elements are then obtained by combining the
measured Pauli correlators according to the Pauli decomposition of the
Hamiltonian,
\begin{equation}
    \langle \psi_0 | \mc{H} U^{k-j} | \psi_0 \rangle = \sum_P \alpha_P \langle \psi_0 | P U^{k-j} | \psi_0 \rangle,
\end{equation}
where $\mc{H} = \sum_P \alpha_P P$. 
The overlap matrix is obtained simply by choosing
$P=I$, giving
$\tilde S_{jk}
=
\langle\psi_0|U^{k-j}|\psi_0\rangle$.
Since the same Hadamard-test circuit is employed for both quantities,
the overlap and Hamiltonian matrices can be evaluated within a unified
measurement framework. Furthermore, commuting Pauli operators may be
measured simultaneously to further reduce the total measurement cost.

\begin{figure}
    \centering
    \includegraphics[width=0.85\linewidth]{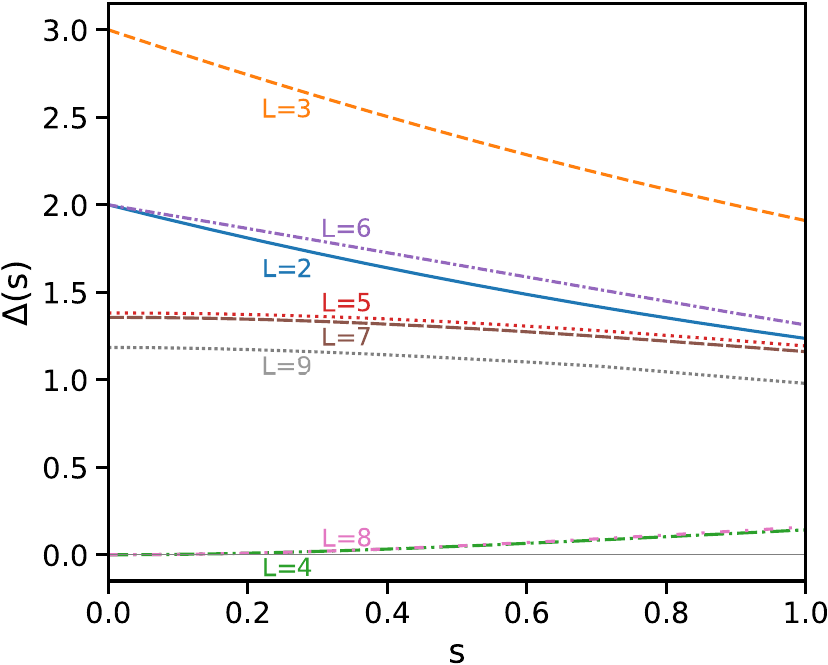}
    \caption{Evolution of the energy gap $\Delta(s)$ between the ground state and the first
        excited state along the linear interpolation $\mc{H}(s)=(1-s)\mc{H}_i+s\mc{H}_f$, 
        where $s=t/T$ is the normalized evolution parameter and $T$ is the total
        adiabatic evolution time [Eq.~\eqref{eq:adia}].
        Results are shown for Hubbard chains with lattice sizes $L=2,3,\ldots,9$.
        } \label{figA2}
\end{figure}

\section{Spectral-gap analysis along the interpolation path}
\label{ap:gap}

To further investigate the origin of the anomalously slow convergence
observed for the $L=4$ and $L=8$ Hubbard chains in
Sec.~\ref{sss:Size-eff}, we analyze the low-energy spectral structure
along a linear interpolation between the non-interacting and
interacting Hamiltonians,
\begin{equation} \label{eq:adia}
    \mathcal{H}(s) = (1-s)\mathcal{H}_i + s\mathcal{H}_f, \qquad 0 \le s \le 1,
\end{equation}
where $\mathcal{H}_i$ denotes the non-interacting Hubbard Hamiltonian
($\mathcal{U}_0=0$) and $\mathcal{H}_f$ is the target interacting
Hamiltonian.
%
%
For each value of $s$, we compute the two lowest eigenvalues by exact
diagonalization and define the instantaneous energy gap as
\begin{equation}
    \Delta(s)=E_1(s)-E_0(s),
\end{equation}
where $E_0(s)$ and $E_1(s)$ are the ground- and first-excited-state
energies, respectively.

Figure~\ref{figA2} shows $\Delta(s)$ for different system sizes.
For most lattice sizes, the energy gap remains finite throughout the
interpolation, whereas the $L=4$ and $L=8$ systems exhibit an
almost vanishing minimum gap.
Although the present QKD algorithm does not employ adiabatic state
preparation, the instantaneous spectral gap provides valuable insight
into the difficulty of resolving the low-energy spectrum. 
When the gap becomes small, neighboring eigenstates acquire relative dynamical
phases more slowly during real-time evolution. 
Consequently, short evolution times generate Krylov basis states that are 
less effective at distinguishing the low-energy manifold, requiring either 
longer evolution times or larger Krylov subspaces to achieve comparable
accuracy.

\begin{figure}
    \centering
    \includegraphics[width=0.85\linewidth]{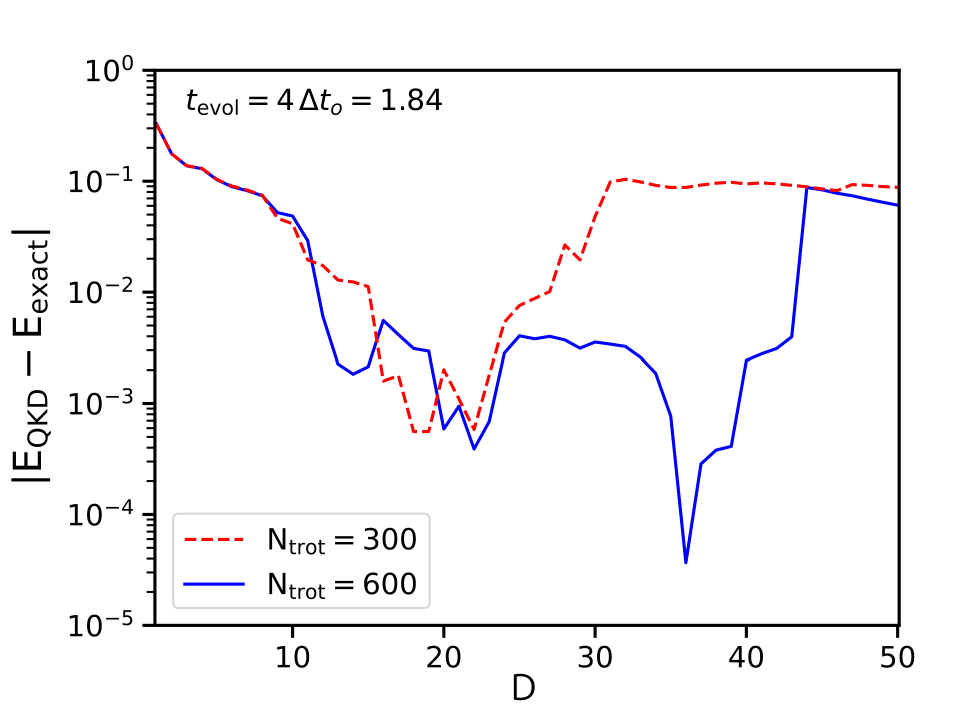}
    \caption{GSE error as a function of the Krylov dimension $D$ for 
        the four-site Hubbard model ($L=4,\ \mc{U}_0=2$) using a longer 
        evolution time $t_\mr{evol} = 4\Delta t_o \approx 1.84$ and $\delta_{\mr{SVT}}=0.9$. 
        Results are shown for $N_\mr{trot}=300$ (red dashed) and $N_\mr{trot}=600$ (blue solid). 
        The improved convergence obtained with the longer evolution time supports the 
        interpretation that systems with small spectral gaps can benefit from longer 
        real-time evolution, provided that the time-evolution operator is implemented 
        with sufficient accuracy.
        } \label{figA3}
\end{figure}

To test this interpretation, we repeated the QKD calculation for the
representative $L=4$ system using a substantially longer evolution
time, $t_\mr{evol}=4\Delta t_o$.
%
The resulting energy error is shown in Fig.~\ref{figA3}. 
Compared with the results presented in
Sec.~\ref{sss:Size-eff}, the longer evolution time substantially
improves the convergence rate and reduces the attainable energy error
before numerical instabilities emerge.
The remaining breakdown at large Krylov dimensions is attributed to accumulated Trotter 
errors, which can be postponed by increasing the number of Trotter steps.

These results support the interpretation that the poor convergence
observed for $L=4$ and $L=8$ originates primarily from their
near-closing spectral gaps rather than from an intrinsic limitation of
the QKD algorithm. At the same time, they illustrate the central
trade-off identified throughout this work: longer evolution times
improve spectral resolution but require a more accurate implementation
of the time-evolution operator to suppress accumulated Trotter errors.



\section{Bootstrap estimation of statistical uncertainty}
\label{ap:bootstrap}

This procedure was used to estimate the error bars reported in
Figs.~\ref{fig8} and \ref{fig9}.
To characterize the statistical uncertainty arising from finite-shot
sampling, we performed $N_\mr{run}=25$ independent realizations of the
complete QKD workflow.
Each realization employed an independent sampling of measurement
outcomes while keeping all algorithmic parameters fixed. For a given Krylov
dimension $D$, the corresponding ground-state energy error was computed as
\begin{equation}
    \Delta E^{(r)}(D)
    =
    \left|
    E_{\mr{QKD}}^{(r)}(D)-E_{\mr{exact}}
    \right|,
\end{equation}
where $r=1,\ldots,N_\mr{run}$ labels the independent realizations.

The statistical uncertainty was estimated using nonparametric bootstrap
resampling. For each Krylov dimension, a bootstrap sample was generated by
randomly drawing $N_\mr{run}$ values with replacement from the set
\begin{equation}
    \left\{
    \Delta E^{(1)},
    \Delta E^{(2)},
    \ldots,
    \Delta E^{(N_\mr{run})}
    \right\},
\end{equation}
and computing the corresponding sample mean. Repeating this procedure
$N_\mr{boot}=1000$ times produced a bootstrap distribution of mean energy
errors,
\begin{equation}
    \left\{
    \overline{\Delta E}^{(1)},
    \overline{\Delta E}^{(2)},
    \ldots,
    \overline{\Delta E}^{(N_\mr{boot})}
    \right\}.
\end{equation}

The reported mean energy error was taken as the average of the bootstrap
distribution,
\begin{equation}
    \mu_{\Delta E}
    =
    \frac{1}{N_\mr{boot}}
    \sum_{b=1}^{N_\mr{boot}}
    \overline{\Delta E}^{(b)},
\end{equation}
while the statistical uncertainty was estimated from the standard deviation
of the bootstrap distribution,
\begin{equation}
    \sigma_{\Delta E}
    =
    \sqrt{
        \frac{1}{N_\mr{boot}-1}
        \sum_{b=1}^{N_\mr{boot}}
        \left(
        \overline{\Delta E}^{(b)}
        -
        \mu_{\Delta E}
        \right)^2
    }.
\end{equation}

Throughout this work, the error bars represent one bootstrap standard
deviation, corresponding to the statistical uncertainty associated with
finite-shot sampling.

\bibliographystyle{apsrev4-2}
\bibliography{references}

\end{document}